\documentclass[11pt]{article}
\usepackage{moriond,epsfig}

\def\Journal#1#2#3#4{{#1} {\bf #2}, #3 (#4)}

\def\be{\begin{equation}}
\def\ee{\end{equation}}
\def\bea{\begin{eqnarray}}
\def\eea{\end{eqnarray}}

\begin{document}
\vspace*{4cm}
\title{The puzzling clustering and bimodality of long GRBs optical afterglow luminosities}

\author{ M. Nardini$^1$, G. Ghisellini$^2$, G. Ghirlanda$^2$ }

\address{$^1$ SISSA/ISAS, Via Beirut 2-4, 34014 Trieste, Italy
\\
$^2$  Osservatorio Astronomico di Brera, via E. Bianchi 46, IÐ23807 Merate, Italy  }

\maketitle\abstracts{
The study of the rest frame properties of long Gamma-Ray bursts (GRBs)
afterglows is a fundamental aspect for a better understanding of the
nature of these powerful explosions. 
The launch of the Swift satellite (November 2004) 
marked
a strong improvement of the
observational capabilities 
of X-rays and optical afterglows.
We studied the intrinsic optical afterglows of a sample of long
GRBs finding  an unexpected clustering and a hint of bimodality of
the optical luminosity distribution (at 12h after the trigger).
Through a Montecarlo simulation we proved that both the observed
clustering and bimodality are not simply due to selection effects but
should hide important information in the understanding of the nature
of the afterglow emission. 
These findings can shed also light on the nature of the 
large fraction of optically dark GRBs.
}

\section{Optical luminosity distribution}

The study of the long GRBs optical luminosity light--curves 
rest frame properties 
in pre-{\it Swift} era 
showed that the luminosity distribution of most
GRBs clusters around a typical value. 
The optical luminosity distribution is much narrower than the 
observed flux distribution and is also narrower than
the distributions of the prompt gamma ray energetics 
(Nardini et al.~\cite{na06}; see also Liang \& Zhang~\cite{lz06}).  
We also found a hint of 
bimodality in the optical luminosity distribution since
3 under luminous GRBs 
were at more than 4 $\sigma$ below the brighter events mean luminosity.

\begin{figure}[h!]
\begin{center}
\psfig{figure=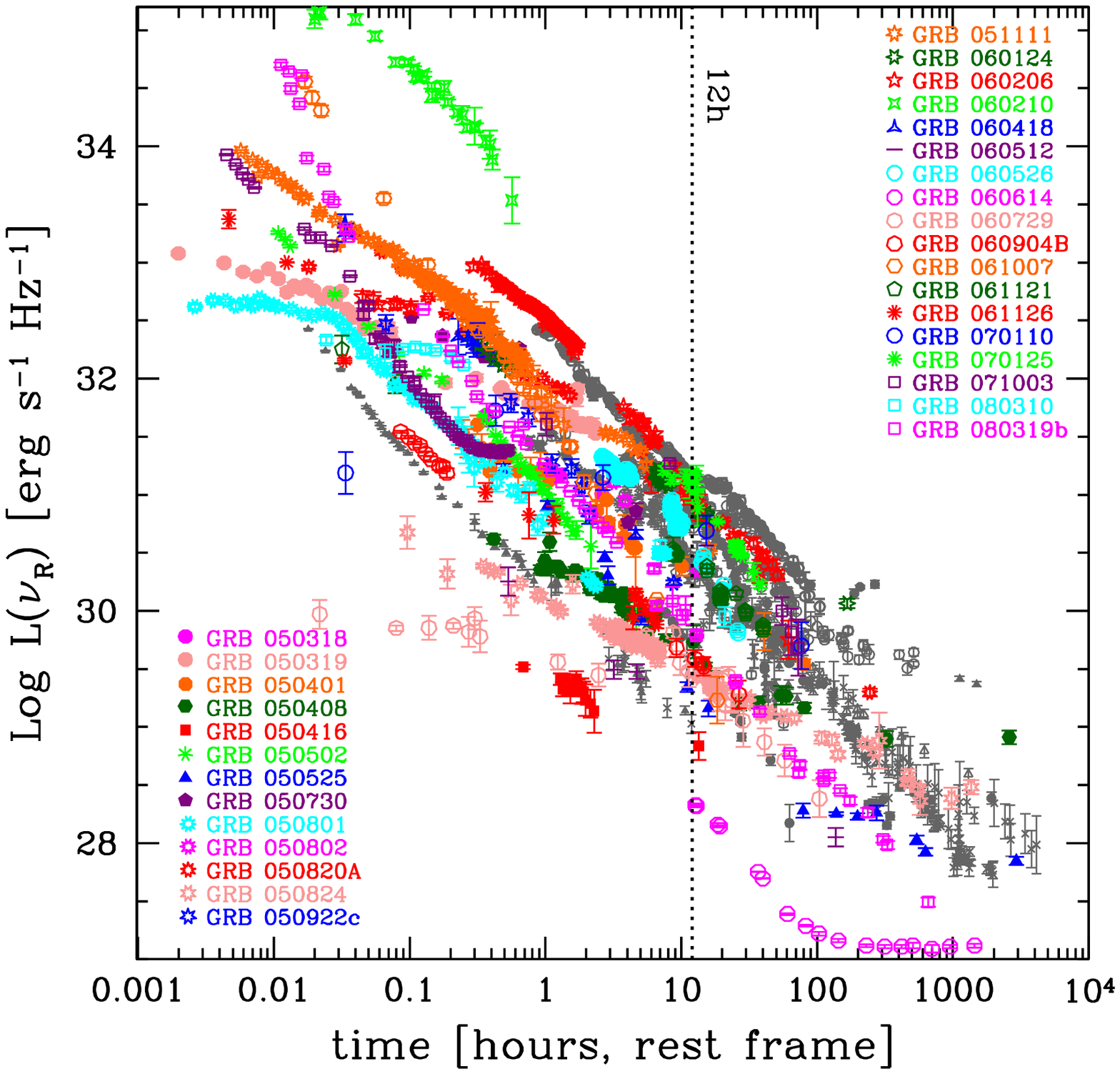,height=85mm,width=7.9cm}
\psfig{figure=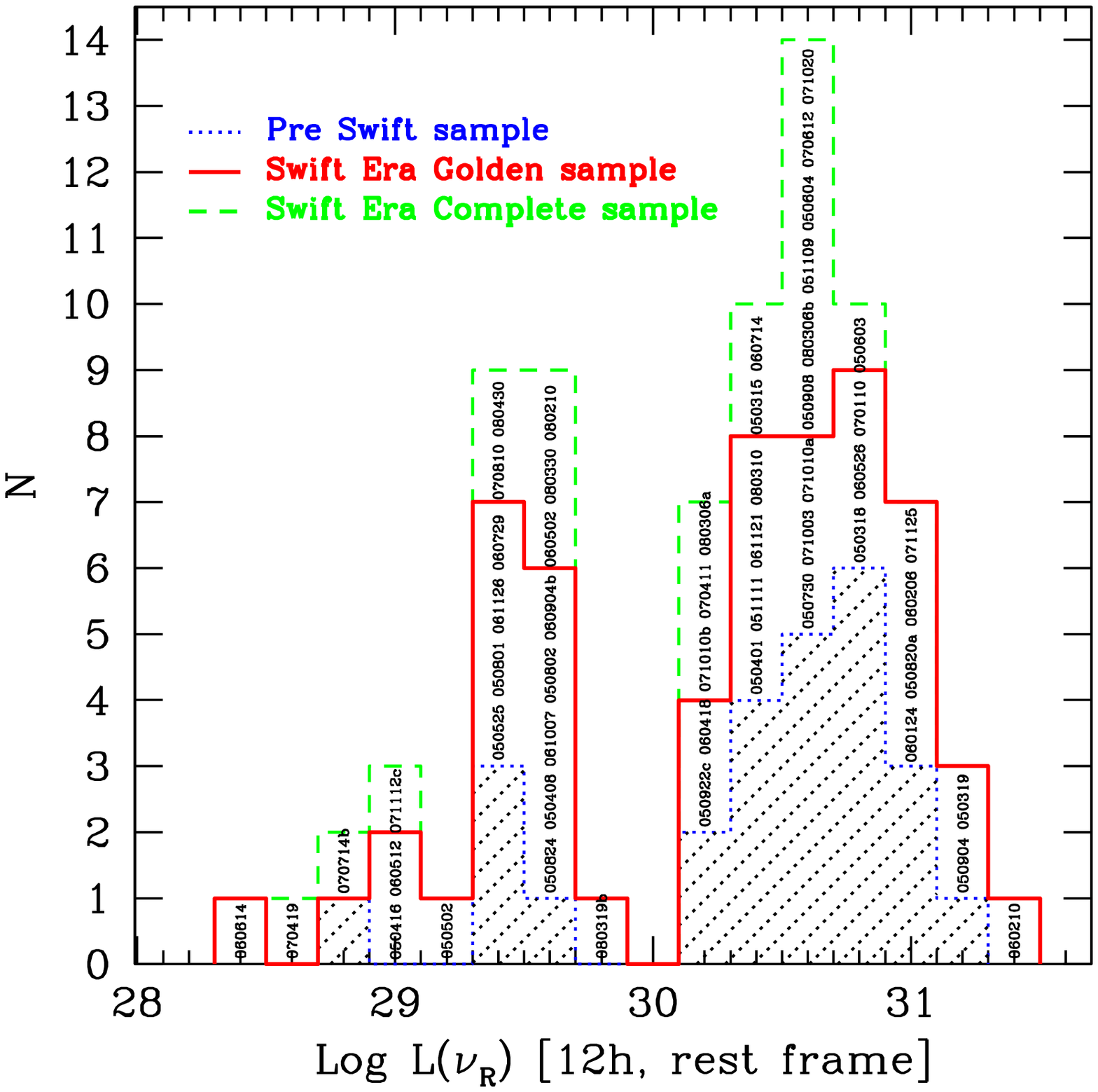 ,height=85mm,width=7.9cm}
\end{center}
\vskip -0.5 cm
\caption{
{\it Left panel}: Rest frame $L_{\nu_{\rm R}}$ light--curves 
of the {\it Swift} GRBs in our sample superposed to the pre-{\it Swift} ones. 
{\it Right panel}: Optical luminosity distribution at 12 h after trigger 
(rest frame time). }
\label{fig1}
\end{figure}

The launch of the  {\it Swift} satellite combined with the availability of 
a net of fast pointing ground based optical telescopes allowed a 
fast and rich optical follow--up starting from a few dozen of seconds after 
the trigger for a large number of events. 
In more recent works we  verified
whether the clustering and the bimodality obtained in the pre--{\it Swift} 
optical luminosity distribution of long GRBs is confirmed by 
more abundant data of the {\it Swift}--era. 
We selected a sample of all the long GRBs 
with known redshift, good optical photometry coverage at some hours after 
the trigger and with a published estimate of the host galaxy dust absorption 
$A_{\rm V}^{\rm host}$. 
As of the end of March 2008 we found a sample of 33 
GRBs fulfilling our selection criteria (golden sample) and 20 without the
$A_{\rm V}^{\rm host}$ estimate. 
In the left panel of Fig. \ref{fig1} 
we plot the rest frame K--corrected and de--reddened monochromatic luminosity 
light--curves at the central frequency of the $R$ band in the rest frame time. 
At very early times (accessible now with {\it Swift}, 
and poorly sampled in pre--{\it Swift} era) the luminosities span some orders 
of magnitudes, but at later times we confirm both the clustering 
and bimodality observed in Nardini et al.~\cite{na06}. 
The right panel of 
Fig. \ref{fig1} shows the luminosity distribution at the common rest frame 
time of 12 h after trigger. 
Note the presence of a gap between the 
under luminous and the brighter events families (for more details see 
Nardini et al.~\cite{na08b}.

\section{The role of selection effects and dark bursts}

In order to test whether the observed clustering and bimodality are 
real or just due to observational biases we applied the method developed 
in Nardini et al.~\cite{na08a}.  
We analysed all the telescopes limiting 
magnitudes of the optically dark GRBs observations and we obtained the 
probability distribution for a GRB to be observed in the $R$ band at 12h 
after trigger with a given limiting magnitude (TSF). 
Through a 
Montecarlo simulation we generated a large number of optical afterglows 
assuming 
i) a redshift distribution following the star formation rate
(Porciani \& Madau~\cite{pm01}), 
ii) different luminosity functions and 
iii) $A_{\rm V}^{\rm host}$ distributions.
We then check if they would be observable using
the limiting magnitudes described by the TSF. 
The luminosity distribution of the observable 
simulated events was then compared with the real one. 

We found that no unimodal intrinsic luminosity function can reproduce the 
real observed distribution.
This can be reproduced either assuming an intrinsically 
bimodal luminosity function or assuming a strong achromatic absorption 
(grey dust) for about half of the simulated afterglows. 
In our 
simulation most of the GRBs belonging to the bright family events are usually 
observable even at large redshifts, while
most of the faint family bursts are below the telescope sensitivity.
The ones we observe are then the tip of the iceberg of a large family 
of optically faint bursts.
The optically dark GRBs can 
therefore belong to these unobservable low luminosity events.

\end{document}